\documentclass[12pt,a4paper]{article}
\usepackage{amsmath}
\usepackage{amssymb}
\usepackage[dvips]{graphicx}

\newcommand{\dd}{\mathrm{d}}
\newcommand{\dbar}{\;\,\bar{}\hspace{-5pt}\dd}

\newcommand{\comments}[1]{}

\title{Measuring the Entropy and Testing \\ the Second Law of Thermodynamics}
\date{March 20, 2002}
\author{Bin Zhou
  \thanks{E-mail: zhoub@ihep.ac.cn}
  \\
   Department of Physics, Beijing Normal University \\
   Beijing 100875, P. R. China \\
   \\
   Institute of High Energy Physics, Academia Sinica \\
   P. O. Box 918-4, Beijing 100039, P. R. China}

\begin{document}
\maketitle

\begin{abstract}

Evidence implies that basic laws of thermodynamics must be tested by
experiments. In this paper, an experiment is designed to measure the entropy
of a system with at least one known (measurable) equation of state, especially
the gas systems. Since the entropy can be measured now, the formulae related to
the second law of thermodynamics can be examined by other experiments. 

\end{abstract}

\section{Introduction}

Among the branches of classical physics, thermodynamics and statistical physics
should be the most fantastic ones, which have attracted so many great minds
to devote their lives and energy to investigating every topic in these fields.
Perhaps among the branches of classical physics, they are the only ones that
leave so many open questions, not only in the field of application, but also
in their foundations.

In the last century, most of the physcists are attracted to the quantized
physics, which can be viewed as the opposite side of classical physics.
Although quantization has been the principal melody of physics ever since then,
thermodynamics and statistical physics are still vivid with their own open
questions. And, even in the so-called modern physics, shadows of thermodynamics
and statistical physics are also seen: In the black hole theory, black holes
are endowed with the temperature and the entropy\cite{blackhole}. Ideas and
concepts of thermodynamics and statistical physics are also applied to the
condensed matter theory and the string theory\cite{string}. In fact, the
history of quantum theory can be traced back to thermodynamics and statistical
physics, where M. Planck use the famous assumption $E=h\nu$ to derive the
spectral energy distribution of black bodies, Planck's formula\cite{Planck}.  
What's more, thermodynamics and statistical physics even paved their way to
theories other than physics, such as Shannon's theory in information theory.
Maybe one day someone cheers that they can be used in economics.

On the one hand, thermodynamics and statistical physics are so widely applied.
On the other hand, there are so many open questions in them --- there are enven 
no universal theories, neither thermodynamic nor statistical, for
non-equilibrium systems. Thus the foundations of thermodynamics, which have
been established for more than one hundred years and which are claimed to be
valid for arbitrary thermal systems and arbitrary processes, could not be
tested by comparing the experimental data with the predictions from a universal
thermodynamics. This sounds not so nice to the society of physics. 

There are other reasons forcing us to seek for the experimental supports of 
the basic laws of thermodynamics. They mainly focus on the easily-misunderstood
concept, heat, in theromdynamics, which has been widely discussed and quarreled
about in literature.

One of such a reason concerns the nature of heat: Is heat a kind of substance
accompanying ordinary matter, as it was said in the old-fashioned caloric
theory, or just the energy transferred from one body to another as the result
of the difference of temperature? As we know, thermodynamics has chosen the
latter. But the following consideration will make thermodynamics to sink into
trouble.

Recall that the Gibbs free energy $G=\mu n$ for an equilibrium system of pure
substance, where $\mu$ is the chemical potential of the substance and $n$ is
its amount meassured in moles. Hence the internal energy $U$ of such a system
reads
\begin{equation}
  U = - pV + TS + \mu n,
\label{Uidentity}
\end{equation}
or something like that. Now let us consider a subsystem of it which has a
virtual boundary that seperates the subsystem from the whole. Now suppose that
the virtual boundary of the subsystem expands slowly such that the center of
mass remains unchanged and the process can be considered as quasi-static. Since
the virtual boundadry and the process are just imaginary, the intensive
quantities $p$, $T$ and $\mu$ are constant while the variations of extensive
quantities $V$, $S$ and $n$ are proportional, that is,
\begin{displaymath}
  \frac{\Delta V}{V_0} = \frac{\Delta S}{S_0} = \frac{\Delta n}{n_0},
\end{displaymath}  
where $V_0$, $S_0$ and $n_0$ are the original volume, entropy and amount of the
subsystem, respectively. Hence, according to the formulism of thermodynamics, 
certain amount of heat $Q=T\ \Delta S\neq 0$ must be absorbed by the subsystem
after its virtual expansion. In other words, as a virtual surface moves in an
equilibrium system, one has to admit that an \emph{additional} amount of heat,
$\dbar Q= T\,\dd S$
has been transferred through it, \emph{accompanying} the substance that goes
through the surface, even though the system is in equilibrium. In this sense,
the caloric theory is somehow restored. On the other hand, however, the zeroth
law of thermodynamics should have repelled such a possibility, because heat
will not be transferred between two systems or two parts of a system provided
there is no difference of temperature.

The other reason that we must \emph{carefully} examine the foundations of
thermodynamics comes from the observation of a transformation in the framework
of the first law of thermodynamics\cite{Zhou}. Briefly speaking, on the one
hand, in order to measure the heat or certain heat-related quantities such as
the heat capacity \emph{accurately}, one must know the internal energy of at
least one substance, the water, for example; On the other hand, according to
the first law of thermodynamics, the knowledge of the internal energy of water,
say, comes from the measurement of heat in various processes of the water.
Obviously, such a chick-and-egg problem can not be solved within the framework
of the first law of thermodynamics. Hence, from the point of view of the first
law of thermodynamics, it is not detectable if the following transformation,
\begin{equation}
  U \longrightarrow U'=U + \phi, \qquad
  \dbar Q \longrightarrow \dbar Q' = \dbar Q + \dd\phi,
\end{equation}
is applied to every thermal system, in which $\phi$ is an arbitrary function of
thermal variables. However, with the second and the third laws of thermodynamics
being considered, the above transformation will be fixed. In other words, in the
point of view of the above transformation, a properly chosen internal energy can
be assigned to each thermal system such that both the second law and the third
law can be satisfied. So one must believe that there are some other methods to
measure the heat in an arbitrary quasi-static process, provided that the second
law and the third law of thermodynamcs hold for every process.

In this paper, such an experiment is proposed. If the theory of thermodynamics
is correct, the entropy of an equilibrium system can be measured. As we have
mensioned in the above, however, the motivation of this paper is not to merely
give a mothod of how to measuring the entropy, hence the heat in a quasi-static
process. The original idea is to test the validity of the whole theory of
thermodynamcs, according to which the experimental data possess the features
that is described in \S\ref{sectcd}. If otherwise, questions of thermodynamics
will arise.

This paper is organized as in the following. In \S\ref{sectionOutline}, we
first labels our ideas of how this experiment is designed. Then follows the
outline of the experiment. Based on the same apparatus, there are two kinds of
methods of how to measure the entropy of an equilibrium system. In
\S\ref{sectionProposalExp}, the formulae are derived in deatails.
In \S\ref{sectcd}, conclusions and further discussions are given.

\section{The Ideas and the Ouline of the Measuring Methods}
\label{sectionOutline}

The ideas of the experiment are described as in the following.

Since, as we have discussed, it remains to be a problem as how to measure the
heat, in order to verify the second law, or strictly speaking, the whole set of
laws, of thermodynamics, we must design an experiment which avoids measuring
quantities related to heat. Hence the heat capacities of various substances
will neither be used as the experimental data nor be measured. Instead, only
quantities such as the volumes, the pressures, the amounts of concerned
substances as well as the temperatures must be measured. This is the basic
point of the proposal experiment.

In deriving the formulae, the formulism of thermodynamics is applied. The
partial derivatives of the entropy or those of the chemical potentials are
replaced of, with the Maxwell's relations being used whenever possible,
because the entropy and the chemical potential cannot be measured
\emph{directly} so far.   
According to the theory of thermodynamics, all the experimental data should
be located on a single line. If one finds this is not the case even after the
errors are considered, there must be some questions in the theory of
thermodynamics.

In measuring the temperatures, the Kelvin temperature scale is needed in
principle. However, in practice this is not so easy to do. This problem is
disposed of in the discussion section, \S\ref{sectcd}.

\begin{figure}[hbt]
\begin{center}
\includegraphics[height=6cm, width=12cm]{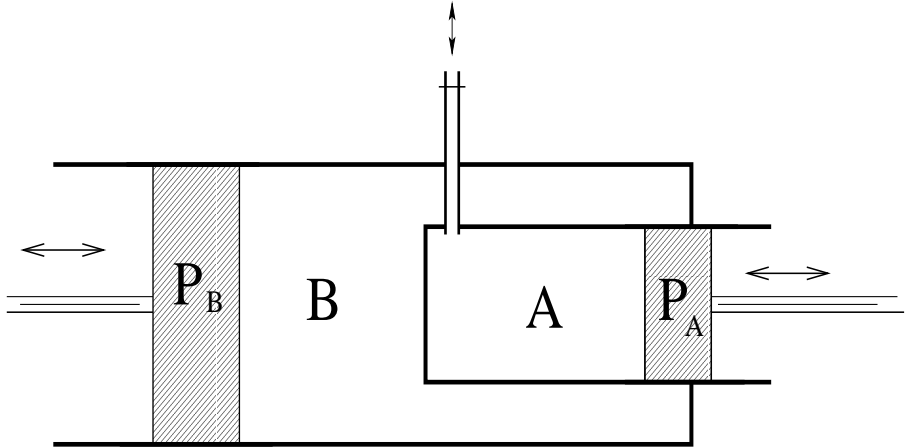}
\caption{Measuring the entropy of gas systems.}
\label{figEntropy}
\end{center}
\end{figure}

The whole system is assumed to be thermally isolated from its surroundings.
This is where most of the errors come from in the measurement. The basic belief
must be accepted in the experiment: Systems will not transfer heat if they have
the same temperature. We can demand the environment of the system to keep the
same temperature as that of the gas in the vessel B (see Figure
\ref{figEntropy}) so that the exchanging heat between the vessel B and the
environment can be ignored. Because the method of thermal isolation is a basic
skill for measurements in thermodynamics, we believe that this will not be a
serious problem.

The experiment is outlined as in the following. 

In the experiment, two kinds of gases, denoted respectively by 1 and 2, are put
into the vessels A and B, respectively, as shown in Figure \ref{figEntropy}.
The vessel A is totally surrounded by the vessel B so that heat transferred out
of the vessel A will be absorbed totally by the gas in the vessel B. The vessel
B is thermally isolated from the environment. Techonologically, one may suppose
that the environment changes its temperature to keep up with the temperature of
the gas in B so that it can be viewed as an adiabatic wall of the the vessel B,
because there will be no heat exchanging between the invironment and the vessel
B. The pistons, P$_\mathrm{A}$ in the vessel A and P$_\mathrm{B}$ in the vessel
B, can move in and out to change the volumes of the gases. In addition, there
is a valve with a pipe inserted into the vessel A, through which small amount
of the gas in the vessel A can be bumped in or out of it. 

The entropies of two different kinds of gases can be measured at the same time.
Let us call the gases 1 and 2, respectively. Suppose that the $i$-th gas is in
equilibrium with the volume $V_i$, the pressure $p_i$, the amount $n_i$ in
the unit of moles and the chemical potential $\mu_i$, where $i=1$ and 2.
Separated by the biathermal walls of the vessel $A$, they share the same
temperature $T$, which is currently measured in the Kelvin temperature scale.

In the first step, gas 1 and gas 2 are put into the vessel A and B,
respectively. the pistons P$_\mathrm{A}$ and P$_\mathrm{B}$ are adjusted so
that the volumes of gas 1 and gas 2 are respectively $V_1$ and $V_2$. Then,
with the pistons fixed, we can always make the gases become thermally
equilibrium with each other at the temperature $T$. After that, we make the
whole system to be thermally isolated from its surroundings.

Now, a small amount $\dd n_1$ of gas 1 is bumped into the vessel A through the
pipe, with the piston P$_\mathrm{A}$ moved slightly meanwhile, making the whole
system out of equilibrium. When the thermal equilibrium is established again,
we can measure again the common temperature, $T + \dd T$, of the gases. Now the
volume of gas 1 is $V_1 + \dd V_1$, and the amount of gas 1 is $n_1 + \dd n_1$.

Then, in the second step, let gas 1 and gas 2 exchange their places, namely,
put gas1 and gas 2 into the vessel B and A, respectively. The pistons are
adjusted so that the volumes of gas 1 and gas 2 are still $V_1$ and $V_2$,
respectively. Similarly, fixing the pistons, we make the gases absorb or emit
certain amount of heat to be in thermal equilibrium with each other at the
temperature $T$ as in the first step. After that, the whole system is again
thermally isolated from its surroundings.

Now, certain amount of gas 2 is bumped into or out of the vessel A. Meanwhile
the piston P$_\mathrm{A}$ is adjusted so that the gases are finally in thermal
equilibrium with each other at the temperature $T + \dd T$. The volume of gas
2 is $V_2 + \dd V_2$, and the amount of gas 2 is $n_2 + \dd n_2$.

In the above steps, a set of data are measured: $\dd V_1/\dd T$ and $\dd n_1/
\dd T$ in the first step, and $\dd V_2/\dd T$ and $\dd n_2/\dd T$ in the second
step. Then the following equation
\begin{eqnarray} 
 & & \frac{1}{n_1}\,\bigg[\,V_1\,\bigg(\frac{\partial p_1}{\partial T}\bigg)
    _{V_1,n_1} - S_1\,\bigg]\,\frac{\dd n_1}{\dd T}
  -\frac{1}{n_2}\,\bigg[\,V_2\,\bigg(\frac{\partial p_2}{\partial T}\bigg)
    _{V_2,n_2} - S_2\,\bigg]\,\frac{\dd n_2}{\dd T}
\nonumber \\
 & = & \bigg(\frac{\partial p_1}{\partial T}\bigg)_{V_1,n_1}\,
    \frac{\dd V_1}{\dd T}
  - \bigg(\frac{\partial p_2}{\partial T}\bigg)_{V_2,n_2}\,\frac{\dd V_2}{\dd T}
  - \frac{\Delta C}{T}
\label{entropyMeasurement}
\end{eqnarray}
should be satisfied, where $S_1 = S_1(T, V_1, n_1)$ and $S_2 = S_2(T, V_2, n_2)$
are the entropies of gas 1 and gas 2, respectively, and $p_1 = p_1(T, V_1, n_1)$
and $p_2 = p_2(T, V_2, n_2)$ are respectively the pressures of gas 1 and gas 2,
while $\Delta C$ is a constant with the dimension of heat capacity, so small
than that of the vessels that it can be ignored, in fact.
In the above equation, the partial derivatives of the pressures with respect to
the temperature must be measured in \emph{other} processes.

With several sets of such data, we can fix the entropies of gas 1 and gas 2.

There is another method to measure the entropy of any one of the gases, $S_1$,
say. Let us put gas 1 into the vessel A and make it to keep equilibrium with
the whole system. Suppose the volume, the amount and the temperature of gas 1
are $V_1$, $n_1$ and $T$, respectively. Again the whole system is thermally
isolated from its surroundings. Then we bump some small amount $\dd n_1$ of
gas 1 into the vessel A. At the same time, the piston A is adjusted so that
the temperature of gas 1 is kept constant throughout the process. If the volume
of gas 1 is now $V_1 + \dd V_1$, then the entropy $S_1$ of gas 1 at the state
$(T, V_1, n_1)$ will be
\begin{equation}
  S_1 = \bigg(\frac{\partial p_1}{\partial T}\bigg)_{V_1,n_1}
    \Big( V_1 - n_1 \frac{\dd V_1}{\dd n_1} \Big).
\label{S1isothermal}
\end{equation}

In the following section, we shall derive the above equation.

\section{The Principles of the Experiment}
\label{sectionProposalExp}

As we know, for a gas system, the first law of thermodynamics can be written in
the form of
\begin{displaymath}
  \dd U = - p \,\dd V + T \,\dd S + \mu \,\dd n.
\end{displaymath}
A Legendre transformation gives the differential of the free energy as
\begin{equation}
  \dd F = - p \,\dd V - S \,\dd T + \mu \,\dd n,
\end{equation}
where the free energy is defined as
$
  F = U - TS.
$
Being a homogeneous function of $V$ and $n$ of order 1 for any given temperature
$T$, the Euler equation for homogeneous functions of order 1 must be satisfied,
in which the temperature is treated as a parameter, giving
\begin{equation}
  F = - pV + \mu n.
\end{equation}
The option form of the above equation can be
$
  U = - pV + TS + \mu n
$
for the internal energy, or\cite{Zemansky}
$
  G = \mu n
$
for the Gibbs free energy.

One of the most important consequences of the above equations is that the
intensive quantities $p$, $T$ and $\mu$ are not functionally independent,
namely, their differentials are constrained by an equation
\begin{equation}
  \dd \mu = \frac{V}{n}\,\dd p - \frac{S}{n}\,\dd T.
\label{diffmu}
\end{equation}
Hence the partial derivative of the chemical potential $\mu$ with
respective to the Kelvin temperature $T$ reads
\begin{equation}
  \bigg(\frac{\partial\mu}{\partial T}\bigg)_{V,n}
  = \frac{V}{n}\bigg(\frac{\partial p}{\partial T}\bigg)_{V,n}
  - \frac{S}{n}.
\end{equation}

By virtue of the above equation, the heat that is absorbed by an open system in
a quasi-static process reads
\begin{equation}
  \dbar Q = T \,\dd S = C_{V,n}\,\dd T
    + T \bigg(\frac{\partial p}{\partial T}\bigg)_{V,n}\dd V
    + \frac{T}{n}\,\bigg[ S - V \bigg(\frac{\partial p}{\partial T}\bigg)_{V,n}
      \bigg]\,\dd n,
\label{dbarQ}
\end{equation} 
in which the partial derivatives of the entropy $S$ with respect to $V$ and $n$,
respectively, have been replaced by the partial derivatives of $p$ and $-\mu$,
both with respective to $T$, according to Maxwell's relations
\begin{equation}
  \bigg(\frac{\partial S}{\partial V}\bigg)_{T,n}
  = \bigg(\frac{\partial p}{\partial T}\bigg)_{V,n},
\qquad
  \bigg(\frac{\partial S}{\partial n}\bigg)_{T,V}
  = - \bigg(\frac{\partial \mu}{\partial T}\bigg)_{V,n}.
\label{MaxwellEqs}
\end{equation}

Now let us focus our attention on the experiment. For the details of
experiment, we refer the readers to \S\ref{sectionOutline}. 

In the first step, gas 1 in A with the pressure $p_1$ and volume $V_1$ is in
thermal equilibrium with gas 2 in B with the pressure $p_2$ and volume $V_2$.
The common temperature of them is $T$. When the volume and the amount of gas 1
are slowly changed, the heat absorbed by gas 1 is, according to
eq.(\ref{dbarQ}),
\begin{displaymath}
  \dbar Q_1 = C_{V,n}^{\ 1}\,\dd T
    + T \bigg(\frac{\partial p_1}{\partial T}\bigg)_{V_1,n_1}\dd V_1
    + \frac{T}{n_1}\,\bigg[ S_1 - V_1 \bigg(\frac{\partial p_1}{\partial T}
      \bigg)_{V_1,n_1} \bigg]\,\dd n_1,
\end{displaymath}
where $C_{V,n}^{\ 1}=C_{V,n}^{\ 1}(T,V_1,n_1)$ is the heat capacity of gas
1 at constant volume $V_1$ and constant amount $n_1$. In this process, since
the volume and the amount of gas 2 is remained to be constant, it absorbs
some heat
\begin{displaymath}
  \dbar Q_2 = C_{V,n}^{\ 2}\,\dd T,
\end{displaymath}
where $C_{V,n}^{\ 2} = C_{V,n}^{\ 2}(T,V_2,n_2)$ is the heat capacity of gas 2
at constant volume $V_2$ and constant amount $n_2$.
Since the whole system is thermally isolated from its surroundings, if the heat
absorbed by the vessels and the pistons is assumed to be
\begin{displaymath}
  \dbar Q_3 = C\,\dd T
\end{displaymath}
where $C$ is the effective heat capacity of the vessels and the pistons, there
must be the equation
\begin{equation}
  \dbar Q_1 + \dbar Q_2 + \dbar Q_3 = 0,
\label{Qeq}
\end{equation}
namely, the heat emitted by gas 1 will be absorbed by gas 2, the
vessels and the pistons, or vice versa. Hence we obtain
\begin{displaymath}
  (C_{V,n}^{\ 1} + C_{V,n}^{\ 2} + C)\,\dd T
   + T\bigg(\frac{\partial p_1}{\partial T}\bigg)_{V_1,n_1}\,\dd V_1
   + \frac{T}{n_1}\,\bigg[ S_1 - V_1 \bigg(\frac{\partial p_1}{\partial T}
      \bigg)_{V_1,n_1} \bigg]\,\dd n_1
  = 0.
\end{displaymath}

From the above equation, we can see that the whole system, when thermally
isolated from its surroundings, has two degrees of freedom. The corresponding
variables may be the volume as well as the amount of gas 1 in vessel A.
Especially, the temperature of the system can be determined by the above two
variables. Thus, for a given state $(T, V_1, n_1)$ of gas 1, one may adjust
the volume and the amount of it so that the temperature keeps invariant. In
such a case, the ratio of the changes of the volume and the amount,
$\frac{\dd V_1}{\dd n_1}$, will be constrained. On the other hand, if this
ratio has been measured, the entropy $S_1$ of gas 1 can be determined as
that in eq.(\ref{S1isothermal}).

Recall that gas 2, the pistons and the vessells won't absorb any heat provided
the temperature remains constant, and that the whole system is thermally
isolated. So the above process is not only isothermal, but also adiabatic. On
the other hand, according to eq.(\ref{dbarQ}), for an open system, if the
process is both isothermal and adiabatic, the entropy of this system can be
determined by the equation having the same form as that in the above, only with
the subscripts being omitted. However, as discussed in the introduction, it
seems that there are some questions in the concept of heat for an open system.
This is not the topic of this paper. We will discuss it in other papers.

Technologically, it is very hard to adjust the volume and the amount of gas 1
so that the temperature keeps constant. Ordinarily, the temperature changes in
the process, which can be referred to as a parameter of the process, yielding
the equation
\begin{displaymath}
  (C_{V,n}^{\ 1} + C_{V,n}^{\ 2} + C)
  + T\bigg(\frac{\partial p_1}{\partial T}\bigg)_{V_1,n_1}\,
    \frac{\dd V_1}{\dd T}
  + \frac{T}{n_1}\,\bigg[ S_1 - V_1 \bigg(\frac{\partial p_1}{\partial T}
      \bigg)_{V_1,n_1} \bigg]\,\frac{\dd n_1}{\dd T}
  = 0.
\end{displaymath}
Noticing that the heat capacities of
gas 1 and gas 2 is totally symmetric in the above equation, one can exchange
their sites and repeat the process, obtaining the following equation:
\begin{displaymath}
  (C_{V,n}^{\ 1} + C_{V,n}^{\ 2} + C')
  + T\bigg(\frac{\partial p_2}{\partial T}\bigg)_{V_2,n_2}\,
    \frac{\dd V_2}{\dd T}
  + \frac{T}{n_2}\,\bigg[ S_2 - V_2 \bigg(\frac{\partial p_2}{\partial T}
      \bigg)_{V_2,n_2} \bigg]\,\frac{\dd n_2}{\dd T}
  = 0.
\end{displaymath}
In the above equation, $C'$ is the affective heat capacity of the pistons and
the vessels when gas 1 is put into the vessel B and 2 into the vessel A.  

With the above two equations,
we expect the result of the experiment to satisfy eq.(\ref{entropyMeasurement}).
In the experiment, we need only to measure the derivatives, with respect to the
Kelvin temperature $T$, of the volume and the amount of the gas that is located
in the vessel A. The coefficients $(\partial p_1/\partial T)_{V_1}$ and
$(\partial p_2/\partial T)_{V_2}$ can and \emph{must} be measured in other
processes. In eq.(\ref{entropyMeasurement}), 
the difference $\Delta C =  C' - C$ of $C$ and $C'$ is so small, compared with
either $C$ or $C'$, that it can be ignored if the accuracy is not highly needed
in the measurement.

In eq.(\ref{entropyMeasurement}), all the quantities except the entropies can
be measured in the experiment. If the theory of thermodynamics is correct, the
entropies of the gases \emph{can} be determined from the data. Having the
method of measuring the
entropy, we can use it to test various statements related to the second law of
thermodynamics, such as the law of increase of entropy\cite{Landau}, the Gibbs'
paradox\cite{Zemansky}. In addition, if both the entropy and the pressure, as
functions of the state of the system, are obtained, the chemical potential can
be integrated --- at least in principle, 
according to eq.(\ref{diffmu}). And one can test whether the Maxwell's relations
such as eqs.(\ref{MaxwellEqs}) is valid for every thermal system, as demanded
by the theory of thermodynamics. Especially, the second equation in
eqs.(\ref{MaxwellEqs}) can be written as
\begin{displaymath}
  V\,\bigg(\frac{\partial p}{\partial T}\bigg)_{V,n}
  = S - n\,\bigg(\frac{\partial S}{\partial n}\bigg)_{T,V}.
\end{displaymath}
It should be verified in the experiments. But, were it verified that $S(T,V,n)$
is a homogeneous function of $V$ and $n$ of order one, the above equation will
be equivalent to the first one in eqs.(\ref{MaxwellEqs}). 

In the above discussion, the entropy of a thermal system of gas can be measured
provided the theory of thermodynamics is correct. As a generalization, we can
conclude that, for a thermal system, if one of its equations of state can be
determined in experiment, then its entropy can be measured by a properly
designed experimental scheme. When its entropy is known, the problem of
measuring heat that it absorbs is solved --- provided the theory of
thermodynamics is correct.

\section{Conclusions and Discussions}
\label{sectcd}

As we have discussed in the above, the entropy of a pure-gas system, which has
three degrees of freedom, can be measured at any given state, $(T, V, n)$, say.
For other systems with three degrees of freedom, its entropy can be similarly
measured provided one of its equations of state can be measured state by state.

Not that the equation of state is not enough to acquire the knowledge of a
thermal system\cite{Callen}. Even for the simpliest case, an ideal gas, the
information of its thermodynamic functions can not be uniquely be determined
by the well-known state of equation, $pV=Nk_BT$. Apart from the general
discussions in the courses, an interesting relativistic case has recently been
discussed in \cite{Pal}. Therefore, the measuring methods for the entropy is
really needed even if the equation of state has been known.

Based on the theory of thermodynamics for quasi-static processes, two methods
are provided to measure the entropy for a gas system at any given state: One is
based on eq.(\ref{entropyMeasurement}), and the other is based on
eq.(\ref{S1isothermal}).

The first method is not so accurate, because there is an unknown quantity
$\Delta C$ which varies from measurement to measurement. Since this is the
difference of the effective heat capacities of the vessels as well as the
pistons, which are considered to remain constant if the accuracy is not needed
to be very high, it is reasonable to be considered as a constant for all the
measurements for a given state. And it is even reasonable to drop it out of
eq.(\ref{entropyMeasurement}), since it is very small.

The second method is more accurate, but it is more difficult to control.
Note that the derivative of $V_1$ with respect to $n_1$ in
eq.(\ref{S1isothermal}) is, in fact, the partial derivative $(\partial V_1/
\partial n_1)_{T,S_1}$. So one can derive that, for the adiabatic isothermal
line $V_1=V_1(T,S_1,n_1)$ where $T$ and $S_1$ are treated as parameters,
\begin{equation}
  \frac{\dd^2 V_1}{\dd n_1^2}
  = - \frac{(\frac{\partial^2 p_1}{\partial T \partial V_1})_{n_1}}
        {(\frac{\partial p_1}{\partial T})_{V_1,n_1}} \,
  \bigg(\frac{\dd V_1}{\dd n_1} - \frac{V_1}{n_1}\bigg)^2.
\end{equation}
Using this relation, one can make the resulted entropy more accurate.

For the first method, which obeys eq.(\ref{entropyMeasurement}), the data
should be located on a 4-dimensional plane spaned by the variables
$\dd n_1/\dd T$, $\dd n_2/\dd T$, $\dd V_1/\dd T$ and $\dd V_2/\dd T$. Of
course, the partial derivatives, $(\partial p_1/\partial T)_{V_1,n_1}$ and
$(\partial p_2/\partial T)_{V_2,n_2}$, must be measured in other experiments.
Although the entropies of two different kinds of gases are measured at the same
time, the resulted entropy of one gas should and must have nothing to with the
other. All the above features are expected according to the theory of
thermodynamics. Were any of them spoiled, questions about the validity of
thermodynamics can be asked.

As for the second method, all the measured data of $\dd V_1/\dd n_1$ should be
consentrated to a definite value, for a given state $(T,V_1,n_1)$ of gas 1.

No matter what method is used, the Kelvin temperature scale must be applied in
order to obtain the correct value of the entropy/entropies. However, in
practice the usual temperature is not the Kelvin temperature. Let the
temperature in the experiment be denoted by $\theta$, say. The temperature $T$
being a increasing function of $\theta$, eq.(\ref{entropyMeasurement}) and
eq.(\ref{S1isothermal}) are turned to be
\begin{eqnarray}
  & &
  \frac{1}{n_1}\,\left[ V_1\bigg(\frac{\partial p_1}{\partial\theta}\bigg)
  _{V_1,n_1} - S_1\,\frac{\dd T}{\dd \theta}\ \right]\,
  \frac{\dd n_1}{\dd \theta}
  -\frac{1}{n_2}\,\left[ V_2\bigg(\frac{\partial p_2}{\partial\theta}\bigg)
  _{V_2,n_2} - S_2\,\frac{\dd T}{\dd \theta}\ \right]\,
  \frac{\dd n_2}{\dd \theta}
\nonumber \\
  & & \ \ \ \ \ \ \ \ \ 
   = \bigg(\frac{\partial p_1}{\partial\theta}\bigg)_{V_1,n_1}
  \frac{\dd V_1}{\dd \theta}
  - \bigg(\frac{\partial p_2}{\partial\theta}\bigg)_{V_2,n_2}
  \frac{\dd V_2}{\dd \theta}
  - \frac{\Delta C}{T}\bigg(\frac{\dd T}{\dd\theta}\bigg)^2,
\\
  & & S_1\,\frac{\dd T}{\dd \theta} = \big(\frac{\partial p_1}{\partial\theta}\bigg)
  _{V_1,n_1}(V_1 - n_1 \frac{\dd V_1}{\dd n_1}),
\end{eqnarray}
respectively. Hence, strictly speaking, we can only measure the quantity
\begin{equation}
  \tilde{S} = S \frac{\dd T}{\dd \theta}
\end{equation}
if the derivative of $T$ with respect to $\theta$ is unknown. 

It is well known that the errors in thermal experiments are very hard to be
evaluated, not only because it is hard to control in the operations, but also
because there is hardly methods, in principle, to eliminate the errors.
Without quantites related to heat, such as heat capacities, being measured,
the accuracy of the measurement methods proposed in this paper is controlable. 
--- At least it is the case in the second method.

As analyzed in the above, the most possible errors come from the following
points:
\begin{itemize}
\item[(1)] In what extent we can set the whole system to be thermally isolated
from its surroundings;
\item[(2)] In what extent the temperature scale leaves from the Kelvin
temperature scale.
\end{itemize}
And, for the first method, it also depends that the quantity $\Delta C$ varies
rather small from measurement to measurement, namely, the effective heat
capacity of the vessels and the pistons remains steadily in various processes
within the first order of variations of the state.

Now that we have the methods of how to measure the entropy of a system with
three degrees of freedom and one equation of state, further experiments can be
designed to test the theory of thermodynamics. For example, one can test
the law of increasing of entropy, and one can test whether the Maxwell
relations such as eqs.(\ref{MaxwellEqs}) hold or not, etc.

\vskip 10mm

\begin{center}
  \textbf{\Large Acknowledgments}
\end{center}

The author wants to thank Prof. H. Y. Guo and Prof. Z. Zhao for helpful
discussions on such topics. He is also thankful to Dr. Xin Wang, Dr. Yu-Guang
Wang, and Dr. Lian-You Shan for stimulating discussions. Special appreciations
are given to Ji-Jun Li, whose effort on thermodynamics several years ago
is a historical backgroud as why the author paid his attention to these
topics.

\comments{
Why is the equation of state not enough?
Kelvin temperature scale
163:
171:
The Gibbs' paradox?
}

\end{document}